\documentclass[usenatbib]{mnras}

\usepackage{newtxtext,newtxmath}

\usepackage[T1]{fontenc}
\usepackage{ae,aecompl}

\usepackage{graphicx}	
\usepackage{amsmath}	
\usepackage{amssymb}	
\usepackage{multirow}
\usepackage{natbib}

\newcommand{\f}{{\cal F}} 

\title[A New Estimate of Galaxy Mass-to-Light]{A New Estimate of Galaxy Mass-to-Light Ratios from Flexion Lensing Statistics}

\author[Fabritius, Goldberg]{
	Joseph M. Fabritius II,$^{1}$\thanks{E-mail: joseph.m.fabritius@drexel.edu}
	David M. Goldberg$^{1}$
	\\
	$^{1}$Department of Physics, Drexel University, Philadelphia, PA 19104\\
}

\date{Accepted XXX. Received YYY; in original form ZZZ}

\pubyear{2022}

\begin{document}
	\label{firstpage}
	\pagerange{\pageref{firstpage}--\pageref{lastpage}}
	\maketitle
	
	\begin{abstract}
		We perform a flexion based weak gravitational analysis of the first two \textit{Hubble Frontier Field} clusters: Abell 2744 and MACS 0416. A parametric method for using radially projected flexion signals as a probe of cluster member mass is described in detail. The normalization and slope of a $L-\theta_{E}$ (as a proxy for $L-\sigma$) scaling relation in each cluster is determined using measured flexion signals. A parallel field analysis is undertaken concurrently to provide a baseline measure of method effectiveness. We find an agreement in the Faber-Jackson slope $\ell$ associated with galaxy age and morphology for both clusters, as well as a theoretical distinction in the cluster normalization mass. 
	\end{abstract}
	
	\begin{keywords}
		gravitational lensing: weak -- galaxies: clusters: general
	\end{keywords}
	
	
	
	\section{INTRODUCTION}\label{sec:Introduction}
	
	Dense galaxy clusters are among the most important test-beds of dark matter dynamics, and over the last two decades extensive gravitational lensing studies have served as outstanding tools for probing them \citep{Metcalf/Madau:2001,Clowe_2006,Munshi2008,Evans/Bridle,Okabe2010,Massey_2015}. While strong gravitational lensing is used to observe multiple lensed images for determining the large-scale mass structure of galaxy clusters \citep{Jauzac2015,Kawamata_2016,Cerny_2018,Parry_2016}, this technique is less suited for probing these systems on smaller scales. Recent developments have shown the combination of galaxy-galaxy lensing and kinematic measurements can help push the utility of strong lensing signals farther than previous work\citep{Grillo2015, Meneghetti2017, Natarajan2017}, modeling galaxy clusters on smaller scales to effectively constrain the substructure. While these developments are greatly improving our understanding of cluster structure and dynamics in the current paradigm \citep{Meneghetti2020, Meneghetti2022}, looking past the strong regime to included higher-order corrections can prove quite fruitful and help compliment future lensing work \citep{Pizzuti2022}. Higher-order weak lensing signals can probe cluster substructure on scales which are not practically detectable through other observational means \citep{Cardone2016,Cain2016,Lanusse2016}. While shear-based weak lensing has been an active field of discovery \citep{Okabe/Umetsu,Umetsu_2014,Medezinski_2016,Finney_2018,Fong2018,Klein2019}, we may also turn to the second-order distortions in a lensed image, also known as flexion, to better understand cluster profiles. 
	
	In this paper we develop a new statistical probe for galaxy clusters using flexion signals. As an initial test of our approach, we present a weak lensing analysis of the underlying mass halos in the Abell 2744 and MACS 0416 clusters. We demonstrate the efficacy of using flexion signals from lensed objects in conjunction with cluster lens properties to constrain the mass-to-light ratio of these cluster profiles. We determine the normalization and slope of the galaxy \textit{L-$\sigma$} Faber-Jackson relation and present the corresponding galaxy velocity dispersion for each cluster.
	
	The outline of the paper is as follows: In Section \ref{sec:flexion} we provide the necessary background of lensing formalism used in our work, a brief overview of the \texttt{Lenser} program used for measuring the flexion signals and discuss our methods for utilizing those signals in a  cluster lensing analysis. In Section \ref{sec:analysis} we discuss the data reduction and image processing used in our work, as well as a description of our mass model and analysis methods. In Section \ref{sec:results} we present the final results of our analysis on the Abell 2744 and MACS 0416 clusters with both a first pass zero prior fit and then a refined analysis merged with a shear cluster mass correction, as well as a summary and discussion of the implications from our results.
	
	\section{Flexion Formalism}\label{sec:flexion}
	
	Using the standard thin lens model, we can relate the convergence in the lens plane, $\kappa$, to a dimensionless potential with $\nabla^{2}\psi = 2 \kappa$. The coordinate mapping problem from foreground, $\vec{\theta}$, to background, $\vec{\beta}$, positions is then related via a second-order expansion in the linear relation of this potential:
	
	\begin{equation}\label{lensTaylor}
		\beta_i = \delta_{ij}\theta^j - \psi_{,ij}\theta^{j} - \frac{1}{2}\psi_{,ijk}\theta^{j}\theta^{k}
	\end{equation}
	
	\noindent where the indices vary over the \textit{x} and \textit{y} cardinal measurements. We can also define the complex derivative operator $\partial = \partial_{1} + i\partial_{2}$. The lensing tensors in the expansion may then be related to the observed lensing effects via
	
	\begin{align*}
		\f =& |\f|e^{i\phi} = \frac{1}{2}\partial\partial^{*}\partial\psi = \partial\kappa\\
	\end{align*}
	
	\noindent showing $\f$ as a gradient, which presents itself through a centroid shift in the lensed object and a useful direct probe of this potential. Measures of third-order foreground potential derivatives ($\psi_{,ijk}$) are thus directly related to the measured flexion signal in a lensed object. 
	
	The combination of a galaxy's characteristic radial size, $a$, and measured flexion produces a scale-invariant dimensionless measure of lensed objects, $|a\vec{\f}|$. Meaning, the same apparent galaxy image produced at different distances will have the same combination of $|a\vec{\f}|$. This then becomes an excellent measure of the \textit{intrinsic flexion} in a distribution of galaxies, with the scatter in that distribution producing a measure of the associated noise in the measured flexion for an object of a given size \citep{Goldberg/Bacon,Goldberg/Leonard,Fabritius1}. This relation, ($\frac{\sigma_{a|\f|}}{a}$), is the form of our measurement uncertainties for subsequent analysis. The intrinsic flexion values are taken from unbiased dimensionless flexion measures in the corresponding parallel fields. 	
	
	\subsection{Lenser}\label{sec:Lenser_Pipeline}
	
	In a previous paper \citep{Fabritius1} the measurement of a flexion signal made use of the developed program  \texttt{Lenser}\footnote{https://github.com/DrexelLenser/Lenser} -- a fast, open-source, minimal-dependency Python tool for estimating lensing signals from real survey data. The unlensed intensity profile of a galaxy can be well described by a particular model with a corresponding set of model parameters \citep{SersicA,Graham/Driver}. A combination of moments-based initial estimation and a modified parameterized ray-tracing known as Ana-
	lytic Image Modeling (AIM)\citep{Flexion/Cain} allows for efficient measurement of lensing signals. 
	
	With initialized parameter estimates provided by the measured image moments, the \texttt{Lenser} pipeline employs a two-step $\chi^2$ minimization:
	\begin{enumerate}
		\item first minimizing over the initially coupled subspace $\{n_s, \theta_s\}$
		\item a final full ten-parameter space local minimization.
	\end{enumerate}
	
	The final parameter space for a fit galaxy is given by:
	
	\[p=\left\{\theta_0^1,\theta_0^2,n_s,\theta_s,q,\phi,\psi,_{111},\psi,_{112},\psi,_{122},\psi,_{222}\right\}.\],
	
	\noindent where $\theta_0^1, \theta_0^2$ are the model centroid locations, $n_s$ is the S\'{e}rsic index, $\theta_s$ is the characteristic radius, $q$ is the semimajor-to-semiminor axis ratio, and $\phi$ is the galaxy orientation. \texttt{Lenser} will also return a direct measure of the galaxy size defined by the quadrupole moments as,
	
	\begin{equation}
		a = \sqrt{(|Q_{11} + Q_{22}|)}
	\end{equation}
	
	\subsection{Flexion Decomposition Technique}\label{sec:flexdecomp}
	
	Weak gravitational lensing has been used in previous studies \citep{WLSharon,McCleary_2015,Medezinski_2016,Okabe_2016} to probe the complex structures of galaxy clusters. Flexion has only been used in a handful of studies \citep{Bird/Goldberg,Cain2016,Leonard2007,Cardone2016} that have made use of non-parametric methods for reconstructing the mass field potential. 
	
	While non-parametric models make no assumptions about the mass-to-light ratios, our study is different. We assume that luminous galaxies correspond to regions of high mass. Under the assumption that lens galaxies lay within the center of Dark Matter halos, using the combination of flexion signal and the lens-plane locations as a proxy for expected substructure centers allows for a unique investigation of galaxy clusters. For simplicity we consider the density profile of an individual lensing galaxy as a Singular Isothermal Sphere (SIS) which gives a flexion signal,
	
	\begin{equation}\label{eq:fsis}
		\f = -\frac{\theta_E}{2 |\theta_L|^2}
	\end{equation}
	
	where $\theta_{E}$ is the Einstein radius and $\theta_L$ is the source-lens angular separation in the plane.

	The choice to utilize an SIS model over the more ubiquitous and mathematically complex Navarro-Frenk-White (NFW) profile is to simplify the direct measurement of a mass tracer ($\theta_E$). The SIS provides a nice mapping between our measured flexion signals, the source-lens separation, and the Einstein radius of the selected lens. In contrast, the more commonly used NFW provides less direct insight\citep{BGRT}. Though these models tend towards an infinite integrated mass, we still have a reasonable scaling relationship between them. Consider the derived power-law,
	
	\begin{equation}
		\mathcal{F} = \frac{n (2-n)}{2}\frac{\left(\theta_E\right)^n}{\left(\theta_L\right)^{n+1}},
	\end{equation}
	
	\noindent which returns a SIS profile for $n$=1. The NFW profile will essentially vary between $n$=0 near the core and $n$=2 at large distances, both of which are not relevant regimes for this analysis. The NFW and SIS profiles return slightly different estimates of the mass (via $\theta_{E}$) but behave consistently.
	
	We can relate the Einstein radius to the  one-dimensional velocity dispersion, $\sigma_v$, of a lensing galaxy with,
	
	\begin{equation}\label{eq:Ein_vel}
		\theta_E = 4 \pi \left(\frac{\sigma_v}{c}\right)^{2} \cdot \frac{D_{LS}}{D_{S}}
	\end{equation} 
	
	\noindent where $D_{LS}$ and $D_{S}$ are the angular diameter distances from lens-to-source and observer-to-source respectively.
	
\begin{figure}
		\centerline{
			\includegraphics[width=\columnwidth]{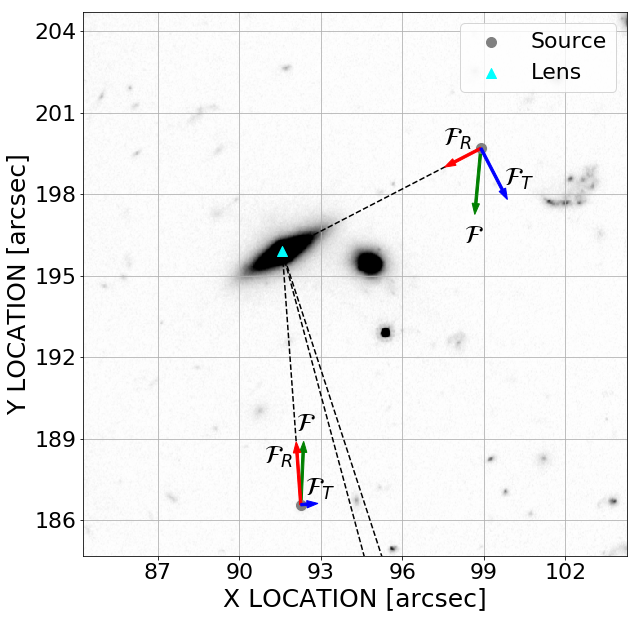}
		}
		\caption{A simple representation of the flexion signal decomposition technique. The cluster member is identified and the measured flexion signal from each candidate source within range is projected onto the straight-line distance in the lens plane (shown as a dashed line). Measured flexion ($\f$), radial flexion ($\f_R$), and tangential flexion($\f_T$) are labeled accordingly. Flexion arrows are drawn proportional to the magnitude for increased visual effect.}
		\label{fig:f_decom_fig}
	\end{figure}

	The detected flexion can be projected toward an \textit{n{th}} nearest-neighbor lensing galaxy. Using the lens-plane straight line distance the unbiased flexion signal is decomposed into two components: a radial ($\f_R$) and tangential ($\f_T$) component, the latter of which should provide an independent measure of noise. A positive radial signal is taken to be oriented inwardly toward a decomposition lens (see Figure \ref{fig:f_decom_fig}).	
	
	The inclusion of the $\f_T$ signal is a key test of this technique. Under the assumption that radial flexion follows the underlying SIS potential directly, it is expected to return a positive signal across the entire cluster field. Conversely, the tangential flexion should favor no bias in direction, and measuring the scatter and mean of this $\f_T$ distribution will help to qualify this.
	
	\section{Data}\label{sec:analysis}
	
	The Hubble Frontier Fields (HFF) is a survey program focused on six prominent Galaxy Clusters which contain large amounts of mass in a compact range \citep{FrontF}. The program uses WFC3 and ACS camera systems  over 70 orbits each to simultaneously image the selected cluster and a nearby parallel field to calibrate the magnification effects of lensed background images. The main focus of the HFF is on deep field observations of faint distant galaxies, which has the added effect of including many bright background objects that can be analyzed for lensing. The subsequent release of the extensive HFF-DeepSpace \citep{HFFDS} multi-wavelength catalogs covering the full breadth of the six Frontier Field clusters allows for a better selection of candidate lens galaxies, cluster member galaxies and foreground object removal. The catalogs contain photometric redshifts, stellar population properties rest-frame magnitudes and colors, and lensing magnification factors, as well as other high-level science data products.
	
	We focus  on the clusters Abell 2744 and MACS 0416, which have been the subject of several previous analysis \citep{Bird/Goldberg,Jauzac2014,Lam/2744,A2744/merger,Owers/2744}. Taking advantage of the high-level science images released to the MAST archives from the HST Frontier Fields Science Products Team, we make use of the Advanced Camera for Surveys (ACS) $I_{814}$, $V_{616}$, and $B_{435}$ filters for color magnitudes with total respective filter integration times corresponding to 46, 14, and 24 orbits. Shape and magnitude measurements are solely taken from $I_{814}$ filter. The Abell 2744 cluster sits at a redshift of $z$ = 0.308 with a field size of roughly 4.5' $\times$ 3.5' in epoch 1 and 3.5’ $\times$ 3.5’ in epoch 2, while MACS 0416 sits at a redshift of $z$ = 0.397 with a field size of 3.5' $\times$ 3.5'.
	
	\subsection{Cluster Member Identification}\label{sec:clusterid}
	
	We first identify candidate lens objects within the relevant cluster field. A key component of the HFF-Deepspace photometry work required the identification and isolation of bright cluster member galaxies. These Bright Cluster Galaxies (BCGs)  form the bulk of the lens object catalogs, and contain associated photometric redshift and $I_{814}$ magnitude measures. Magnitude measures were originally reported in AB and converted to absolute magnitude for our analysis. The publically available lens catalogs for Abell 2744 and MACS 0416 \citep{MACS0416/stronglens} are position matched with the HFF-Deepspace objects to create a larger cluster catalog. Redshift bounds were further placed on the cluster candidates for improved accuracy, with the Abell 2744 cluster defined as 0.308 $\pm$ 0.05 and MACS 0416 defined within the range of 0.382 $\pm$ 0.014 \citep{bergamini2019}.
	
	For the parallel fields, which in principal contain no actual cluster member objects, we created phantom locations corresponding to identified BCGs in the field in order to execute a similar flexion decomposition analysis. Objects to be used in place of cluster members were identified using the same redshift windows used to define the corresponding cluster plane windows. While the lens plane windows defined are much more sparsely populated than the cluster fields, the identified objects in the dummy lens-plane are still sufficient for a parallel field decomposition signal analysis. The use of these decomposed parallel field signals in conjunction with the tangential flexion signals within the actual cluster fields provide a broad cover of a null signal check.

	\subsection{Image Preprocessing}\label{sec:p2_imgpp}
	
	\texttt{Lenser} requires an input galaxy image to estimate and subtract a background, and estimate the sky and Poisson noise to use as a noise map weighting. A mask is then added so as to include only relevant pixels in the input image, reducing error from spurious light sources. A secondary measure for reducing the inclusion of non-galaxy object pixels is to include a segmentation map to identify likely outliers. This method sets pixels in the weight map to zero if they are associated with other objects in the segmentation map or located closer to another object in the stamp, a so-called \"{u}berseg technique that has been shown to greatly improve the model fitting on deblended objects \citep{DES}.
	
	All initial mosaic images are cropped to exclude regions of poor coverage in order to reduce the background estimation bias of the extraction program. A first pass through the open-source \texttt{Source} \texttt{Extractor} program \citep{SExtractor} is used to create a full mosaic segment map of detected bright objects with a minimum of 25 pixels above a 1$\sigma$ threshold. The locations of identified cluster members are position matched to the segment map and filled with randomly sampled noise drawn from the corresponding mosaic noise map set to the mean of the local background. If available, an exposure map was also included in the extraction to improve object detection. This is a modification of the hot-cold search method used in most lensing models \citep{RixHC,LeonardHC}.
	
	The noise masked output image is then run through a second pass of \texttt{Source} \texttt{Extractor} run in dual-image mode across all available bands, targeting the expected source population with a minimum of 15 pixels above a 2$\sigma$ threshold. After all objects were identified across the three relevant bands, a selection criteria was applied to exclude source contaminants or low S/N objects. Identified objects outside of the magnitude range 20 $<$ $mag_{F814W}$ $<$ 28, (2.0 $\times$ FWHM$_{PSF})$ $<$ FWHM $<$ 0.9'' or S/N $<$ 20 were rejected. Additionally, the \texttt{Source} \texttt{Extractor} internal flagging was used to reject any object identified as incomplete (FLAG $\geq$ 8). Square postage stamps for the selected sources were generated using the centroid location with a windowed radial extent set to 4.5 times the calculated half-light radii, in order to include the light at the edges of an object important for flexion fitting.
	
	For the Abell 2744 cluster in particular, the presence of two available field observations motivated the need to apply the full process to each individually. If an object was present in both \texttt{Source} \texttt{Extractor} runs, the secondary epoch object was favored as the image had a richer exposure. In order to remove any spurious detections at the edge of the field-of-view, a 10 arcsecond interior buffer was used, excluding any \texttt{SExtracted} sources in this `no-mans-land'. 
	
	The final step in image preprocessing was the implementation of a third pass for the individual stamps. This final pass was run with 'hot' parameters in order to create detailed segment maps that would identify any spurious objects in the stamp that are not part of the centered source. With the identified image stamps collated for \texttt{Lenser} analysis, a PSF stamp is also included. The finalized stamp images are detected in the $I_{814W}$ band. For the accompanying PSF convolution we take the associated PSF stamp from the DeepSpace photometry data release \citep{HFFDS}.

	\subsection{Mass Model}\label{sec:massmod}
	
	In order to probe the underlying mass structure of the galaxy clusters, we assume a scaling model of proportionality between the ensemble mass via a measure of the Einstein radius and the luminosity of cluster members:
	
	\begin{equation}\label{eq:massmod}
		\frac{\theta_E}{\theta_{E}^{\star}} = \left(\frac{L}{L^{\star}}\right)^\ell = \left(10^{\left(0.4\cdot\ell\left(M^{\star} - M_i\right)\right)}\right).
	\end{equation}

	Using an assumed SIS model for the underlying signal of the radial flexion measures, equations \ref{eq:fsis} and \ref{eq:massmod} allow us to probe the relationship between $\theta_{E}^{\star}$ and $\ell$. This can be seen as a modified probe of the Faber-Jackson relation, using Equation \ref{eq:Ein_vel}. The assumed SIS lens produces its own flexion fields, adding to any external background lensing field in the local area. In the case of our target source exhibiting a flexion signal inconsistent with the overall smoothed flexion fields, a local mass should be responsible for the extra observed signal. Thus, the measured flexion signal from a lensed source object can be used to quantify the mass of an individual lens galaxy via Equation \ref{eq:fsis}. Note, we considered the possibility of including a correction for a smoothed cluster contribution but the contribution in aggregate is smaller than the uncertainty in galaxy shape variance. We also stress that the model makes no assumptions about source-lens pairs prior to analysis, and the uncorrelated orientations of source objects and lens members should absolve individually affected contributions that would be closer to the main cluster core.
	
	Thus, our model uses source-lens distance in the plane, a source flexion signal and the corresponding luminosity measure of the lens galaxy to produce a fit for the normalized Einstein radius and mass-to-light slope of the ensemble distribution of cluster member galaxies. We index the cluster member mass to the brightness. The reason for utilizing a $L-\theta_{E}^{\star}$ relation in this study, as opposed to the more conventional $L-\sigma$ notation is to preserve the mass measure as an unbiased metric. While $\sigma_v$ is necessarily a positive value, $\theta_{E}^{\star}$ may freely vary between positive and negative for a cluster lens measurement. A returned negative or near-zero value should indicate a non-useful decomposed signal, and a lack of detectable substructure in the relevant field. 
	
	We may also merge the flexion-only based results with shear estimates of the smoothed background potential from the cluster. This applies a correction to the flexion and to the underlying convergence via the equation
	
	\begin{equation}
		\mathcal{F} = f \left(1 - \kappa\right)
	\end{equation}
	
	\noindent where $f$ is the \textit{reduced flexion} signal reported by the \texttt{Lenser} analysis tool. We make use of the publicly available convergence maps from the CATs (Clusters As Telescopes)	\footnote{https://archive.stsci.edu/pub/hlsp/frontier/macs0416/models/cats/v4.1/} team to apply the measured $\kappa$ signal at the measured point. 
	
	Cluster environments may affect the lensing distortions exhibited by a lensed galaxy image. We also include a cluster mass correction to our analysis by considering the contribution to any flexion signal from the underlying cluster potential. We achieve this by subtracting additional projected SIS signals for identified large dark-matter clusters in the relevant cluster fields. Previous studies\citep{Jauzac2014, Jauzac2015} found that for both Abell 2744 and MACS 0416 there were two large dark-matter clusters measured from the shear fields, with associated parameters shown in Table \ref{tab:dm}. We introduce a SIS flexion signal for both component masses in the relevant cluster at the specified location using an Einstein radius proportional to the derived mass. These cluster signals are similarly decomposed along the $\theta_L$ vector of a source-lens pair and then subtracted from the final reported radial/tangential value used in the model fit. 
	
	\section{Results}\label{sec:results}
	
	After the source objects stamps were prepared, the relevant stamps were processed with the \texttt{Lenser} program. Data reduction was a simple cut in of any lensing fit model which returned multiple images within the model stamp frame, as this is indicative of a poor fit. A minimum size cut-off of 7 pixels was also implemented to select for the most useful signals. For Abell 2744, the routine returned an initial 642 source objects, and 128 lens objects in the cluster field. The sparser parallel field contained 491 sources and 27 lens locations. For MACS 0416 cluster field, there were an initial 559 analyzed sources and 159 identified lenses. The corresponding parallel field contained 297 sources and 12 lens locations. Because of the overlap in cluster member positions, individual source flexions can contribute multiple decomposed signals. This effectively increases the utility for the number of objects our analysis. 
	
	With a final catalog of processed objects and field associated candidate cluster members, our model fitting was applied to the four analyzed fields (two clusters and two parallel fields). An iterative process of identifying a lens object and all sources within a 30'' cut-off radius was employed. For each source inside this select region, the flexion signal was decomposed into components $\f_R$ and $\f_T$ toward the selected lens. With the source-to-lens distance, $\theta_L$, these signals were independently used to measure the mass-model.
	
\begin{figure}
		\centerline{
			\includegraphics[width=\columnwidth]{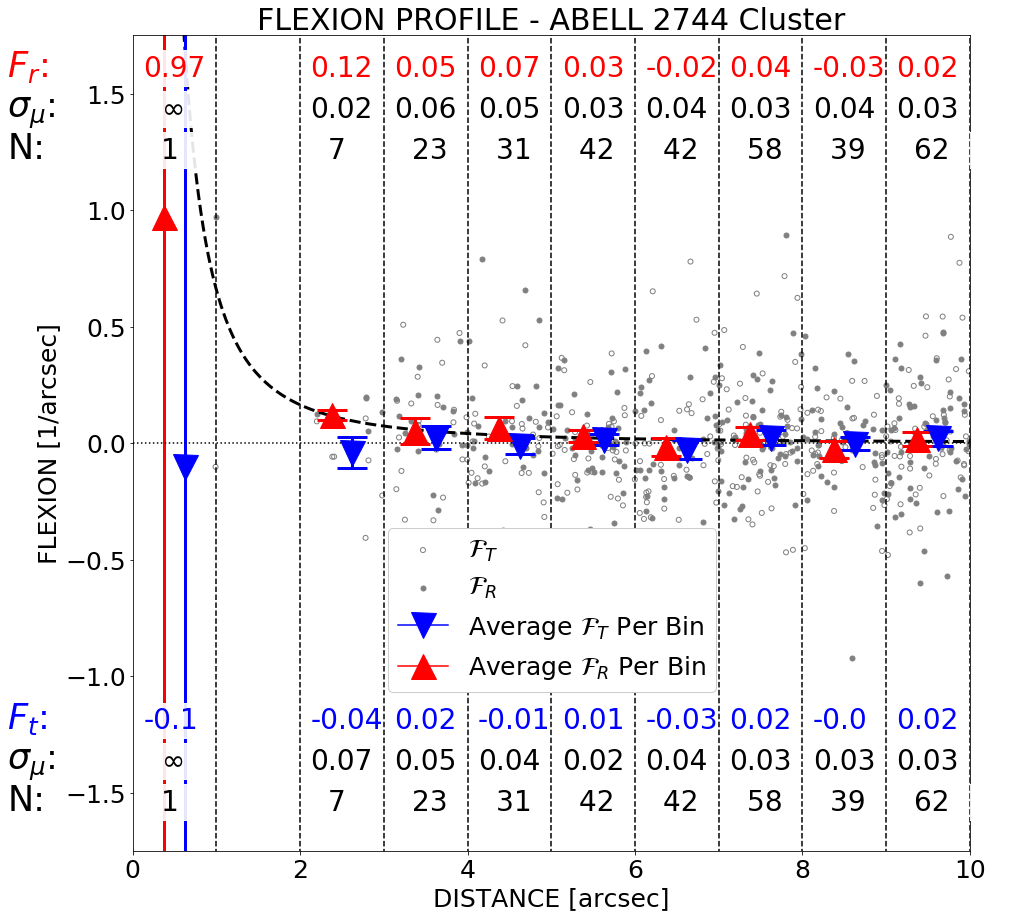}
		}
		\caption{The radial profile for the Abell 2744 cluster signals. Radial distance is reported in 1'' bins. An SIS fit for the radial flexion components is overlaid.}
		\label{fig:a2744_prof}
	\end{figure}

\begin{figure}
		\centerline{
			\includegraphics[width=\columnwidth]{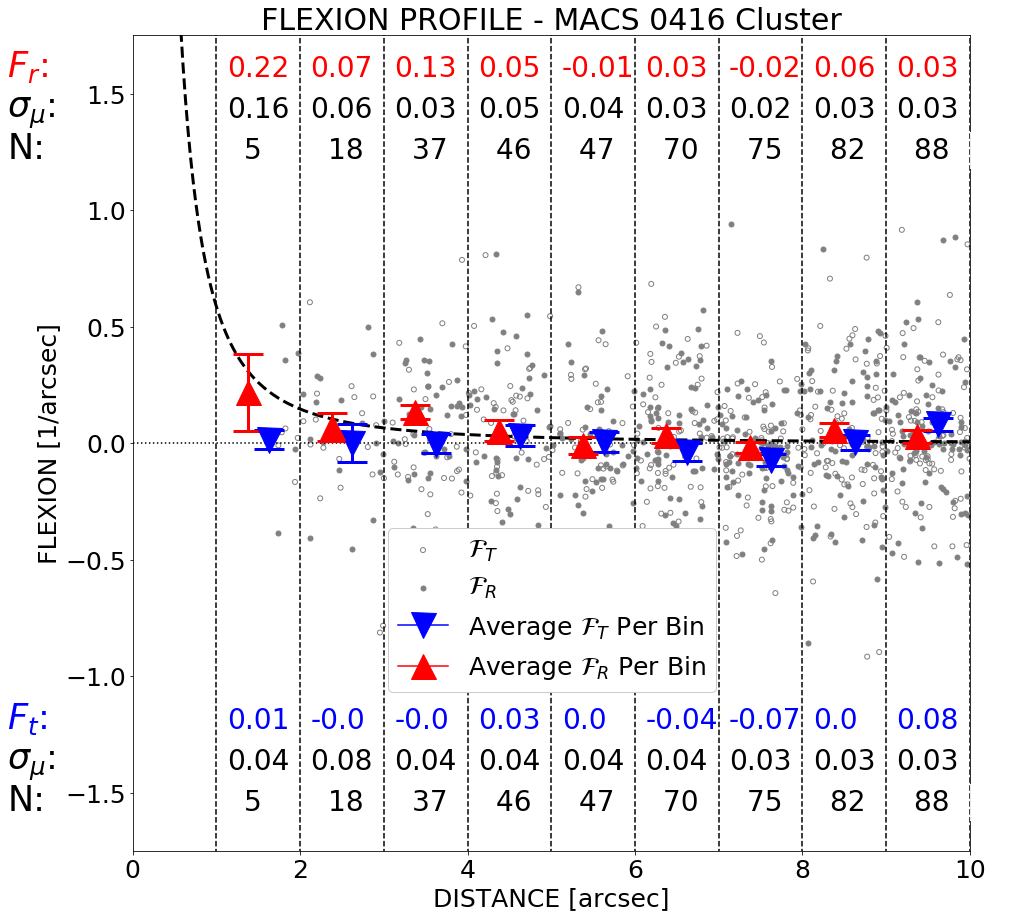}
		}
		\caption{The radial profile for the MACS 0416 cluster signals. Radial distance is reported in 1'' bins. An SIS fit for the radial flexion components is overlaid.}
		\label{fig:m0416_prof}
	\end{figure}

	For the model analysis, we generate two estimates of the substructure masses.  In the first, we suppose that we know nothing about the cluster, in which case we assume a zero prior and no correction associated with the underlying convergence. The flexion profiles for the Abell 2744 cluster and MACS 0416 cluster can be seen in Figures \ref{fig:a2744_prof} and \ref{fig:m0416_prof}. Within a 1'' bin, the radial and tangential signals are averaged with an associated weighted average uncertainty. The values for $\f_R$ and $\f_T$ can be seen to broadly follow two trends: A positively favored signal for the radial component and a tangential signal consistent with zero. While this behavior converges around the 6'' mark, there is still functional information to be gained from the signals at greater distances.
	
\begin{figure}
		\centerline{
			\includegraphics[width=\columnwidth]{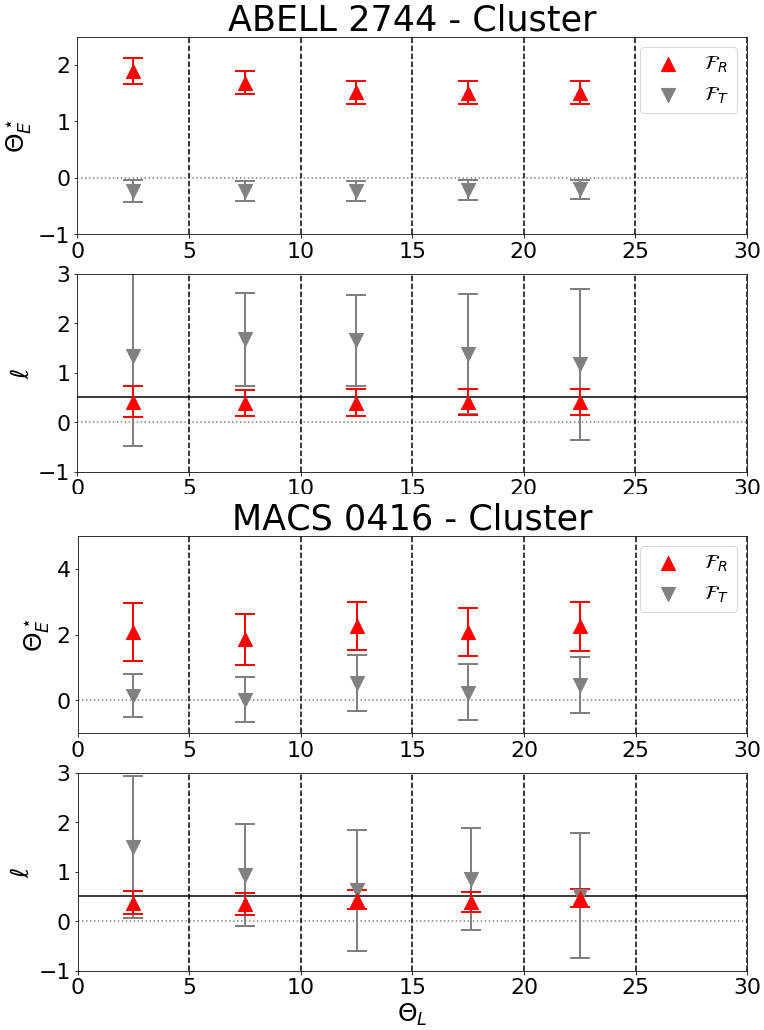}
		}
		\caption{The two-parameter mass model fit from radial flexion of increasing inclusion radius for both the Abell 2744 (top) and MACS 0416 clusters (bottom). The parameter fit values and corresponding uncertainties are taken from a \texttt{scipy.optimize} minimization scheme, selecting all decomposed source signals within inclusive annuli at increments of 5''. The solid line on the $\ell$ fit plots is a visualization for the expected value of $\ell$ = 0.5}
		\label{fig:mass_mod_fit}
	\end{figure}

	To investigate how this two-parameter fit varies with increasing radial inclusion, we can expand this inclusion radius. The decomposed sources interior to the chosen distance bin are used to fit the relevant parameters of Equation \ref{eq:massmod}. The parameter values with associated uncertainties for the first cluster field can be seen in Figures \ref{fig:mass_mod_fit}. It can be seen that the $\theta_{E}^{\star}$ value quickly converges for both the $\f_R$ and $\f_T$ signal fits. Importantly, as this is a trace measure of the underlying cluster mass, the tangential signal is consistently returning a near-zero fit while the positive radial signals are definitively measuring an disturbance in the background potential.The mass-to-light slope measure $\ell$ also converges rather quickly to a value around 0.5 for the radial measurements. From this we can see that any signals outside of 15'' are not likely to be contributing strongly to the cluster parameter fit, and so the final values are all reported from sources within 15'' of a cluster member. 
	
	\begin{table}
		\begin{tabular}{|c|cc|cc|}
			\hline
			Cluster   & \multicolumn{2}{|c|}{Abell 2744}   & \multicolumn{2}{c|}{MACS 0416}    \\ \hline
			$\alpha$ (deg)  & \multicolumn{2}{|c|}{3.586259}    & \multicolumn{2}{c|}{64.0381013}   \\ \hline
			$\delta$ (deg)  & \multicolumn{2}{|c|}{-30.400174}   & \multicolumn{2}{c|}{-24.0674860}  \\ \hline
			Component & \multicolumn{1}{|c|}{$\#$1}  & $\#$2   & \multicolumn{1}{c|}{\#1}  & \#2   \\ \hline
			$\Delta_{RA}$        & \multicolumn{1}{|c|}{–4.9} & –15.7 & \multicolumn{1}{|c|}{–4.5} & 24.5  \\ \hline
			$\Delta_{DEC}$       & \multicolumn{1}{|c|}{2.7}  & –17.2 & \multicolumn{1}{c|}{1.5}  & –44.5 \\ \hline
			$\sigma$     & \multicolumn{1}{c|}{1263} & 134   & \multicolumn{1}{c|}{779}  & 995   \\ \hline
		\end{tabular}
		\caption{The major dark-matter cluster parameters used for cluster correction measurements, taken from \citep{Jauzac2014,Jauzac2015}. Coordinates are quoted in arcseconds with respect to $\alpha$ and $\delta$ for the reported cluster.}
		\label{tab:dm}
	\end{table}
	
\begin{figure}
		\centerline{
			\includegraphics[width=\columnwidth]{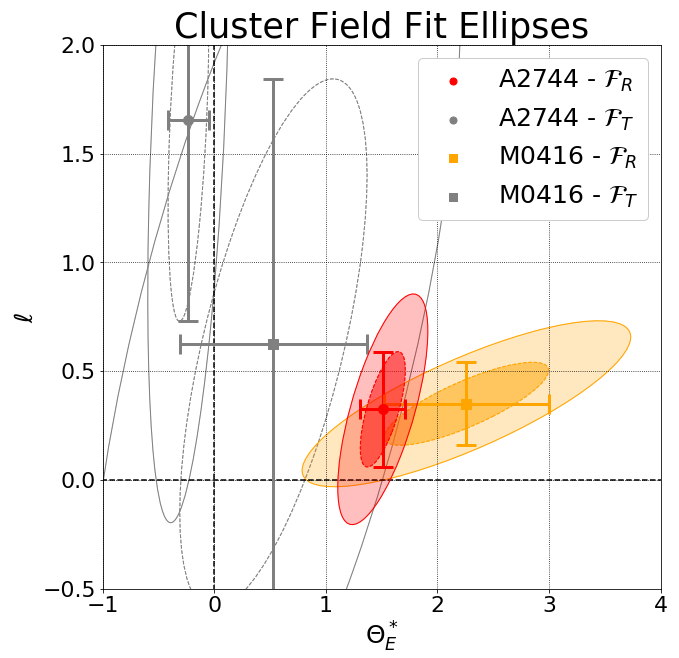}
		}
		\caption{Error ellipses for the fit radial flexion signal in the Abell 2744 and MACS 0416 cluster fields. Cluster field measurements derived from the radial decomposed signals are shown in solid colors (Abell 2744 in red, MACS 0416 in orange). Both fits are mutiple $\sigma$ signals, indicating a direct measure of the lens galaxy mass from our weak-lensing analysis. The tangential signal fits are shown in gray, with a much larger range in errors but consistent with a null signal.}
		\label{fig:comb_clu_fits}
	\end{figure}

	The final results for all fields are presented in the associated Table \ref{tab:results2}. While the cluster corrected signal model returns slightly lower values in the $\theta_{E}^{*}$, the parameter fits are notably more precise, returning reduced uncertainties and more robustly supporting a positive signal detection. The individual two parameter error ellipses for the two analyzed cluster fields are shown in Figure \ref{fig:comb_clu_fits}. Across all fields, only the radially derived signals return persistent non-zero  parameter values, while tangential fits and all parallel signals produce parameter fit values consistent with zero. From the error ellipse plots, it can be seen that only the radial cluster flexion signals have positive correlations clearly separated from a zero-fit in the measured normalization mass. 
	
\begin{figure}
		\centerline{
			\includegraphics[width=3.5in]{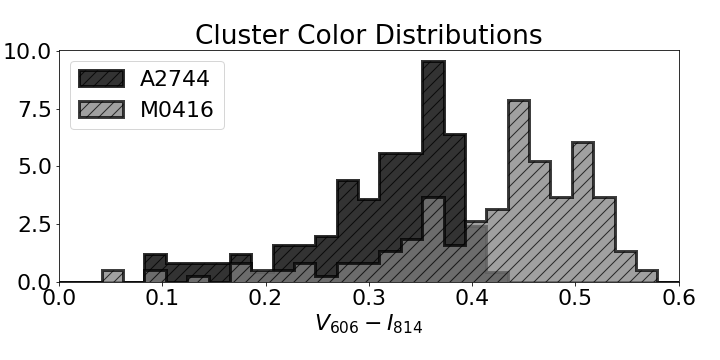}
		}
		\caption{The $V_{606}-I_{814}$ color distributions for both target clusters. A clear median separation can be seen in this filter.}
		\label{fig:color}
	\end{figure}

	The measure of the Einstein radius is a proxy for mass/velocity dispersion, Equation \ref{eq:Ein_vel}, and thus $\theta_E^\star$ is a measure of the cluster normalization mass. Looking at the available color bands there is a clear distinction between the two target clusters (Figure \ref{fig:color}). Abell 2744 has a median value (0.33) lower than MACS 0416 (0.45), which is indicative of a younger cluster. A broad description would anticipate a lower expected mass-to-light ratio for a bluer cluster.  We see from the final analysis in Table \ref{tab:results2} that this broad behavior is confirmed, with the mass-to-light normalization roughly twice as high in MACS 0416 than Abell 2744. 
	
	\begin{table*}
		\begin{tabular}{|c|c|c|c|c|c|c|c|c|c|}
			\hline
			Field                       & Region                                                                    & Signal                          & $\theta_{E_{SIS}}$ {[}arcsec{]} & $\theta_E^*$ {[}arcsec{]} & $\sigma_v^*$ {[}km/s{]} & $\ell$           & $N_{lens}$          & $N_{signals}$        & $N_{sources}$       \\ \hline
			\multirow{6}{*}{Abell 2744} & \multirow{2}{*}{\begin{tabular}[c]{@{}c@{}}Cluster\\ (Null)\end{tabular}} & $\f_R$ & 1.86 $\pm$ 0.27                       & 2.25 $\pm$ 0.46                             & 302 $\pm$ 31                              & 0.62 $\pm$ 0.57  & \multirow{4}{*}{128} & \multirow{4}{*}{628} & \multirow{4}{*}{280} \\ \cline{3-7}
			&                                                                           & $\f_T$ & -0.17 $\pm$ 0.24                      & -0.1 $\pm$ 0.15                             & -                                                      & -0.94 $\pm$ 0.44 &                      &                      &                      \\ \cline{2-7}
			& \multirow{2}{*}{Cluster (Correction)}                                     & $\f_R$ & \textbf{1.31} $\pm$ \textbf{0.13}                       & \textbf{1.52} $\pm$ \textbf{0.04}                             & \textbf{268} $\pm$ \textbf{4}                               & \textbf{0.42} $\pm$ \textbf{0.07}  &                      &                      &                      \\ \cline{3-7}
			&                                                                           & $\f_T$ & -0.13 $\pm$ 0.12                      & -0.23 $\pm$ 0.03                            & -                                                      & 1.75 $\pm$ 0.86  &                      &                      &                      \\ \cline{2-10} 
			& \multirow{2}{*}{Parallel}                                                 & $\f_R$ & -0.09 $\pm$ 1.39                      & 0.54 $\pm$ 7.29                             & -                                                      & 5.36 $\pm$ 26.47 & \multirow{2}{*}{27}  & \multirow{2}{*}{81}  & \multirow{2}{*}{67}  \\ \cline{3-7}
			&                                                                           & $\f_T$ & -1.72 $\pm$ 1.38                      & -2.61 $\pm$ 3.13                            & -                                                      & 0.35 $\pm$ 1.13  &                      &                      &                      \\ \hline
			\multirow{6}{*}{MACS 0416}  & \multirow{2}{*}{Cluster (Null)}                                           & $\f_R$ & 1.37 $\pm$ 0.25                       & 4.55 $\pm$ 1.22                             & 531 $\pm$ 72                              & 0.49 $\pm$ 0.15  & \multirow{4}{*}{159} & \multirow{4}{*}{584} & \multirow{4}{*}{254} \\ \cline{3-7}
			&                                                                           & $\f_T$ & 0.0002 $\pm$ 0.27                     & -0.47 $\pm$ 1.58                            & -                                                      & 0.63 $\pm$ 2.01  &                      &                      &                      \\ \cline{2-7}
			& \multirow{2}{*}{Cluster (Correction)}                                     & $\f_R$ & \textbf{1.17} $\pm$ \textbf{0.24 }                      & \textbf{2.26} $\pm$ \textbf{0.43}                             & \textbf{327} $\pm$ \textbf{30}                              & \textbf{0.53} $\pm$ \textbf{0.70}  &                      &                      &                      \\ \cline{3-7}
			&                                                                           & $\f_T$ & 0.17 $\pm$ 0.25                       & 0.53 $\pm$ 0.70                             & -                                                      & 0.62 $\pm$ 1.49  &                      &                      &                      \\ \cline{2-10} 
			& \multirow{2}{*}{Parallel}                                                 & $\f_R$ & -4.82 $\pm$ 5.17                      & 0.57 $\pm$ 5.60                             & -                                                      & -0.49 $\pm$ 1.35 & \multirow{2}{*}{12}  & \multirow{2}{*}{51}  & \multirow{2}{*}{48}  \\ \cline{3-7}
			&                                                                           & $\f_T$ & 0.11 $\pm$ 4.72                       & -9.25 $\pm$ 7.96                            & -                                                      & 1.54 $\pm$ 3.99  &                      &                      &                      \\ \hline
		\end{tabular}
		\caption{Final model fit results for all four analyzed fields, including convergence and cluster corrections. Values are returned inside a cut-off radius of $\theta_L$ $\leq$ 15''. Cluster radial signal fit parameters are bolded for clarity.}
		\label{tab:results2}
	\end{table*}
	
	We can compare the final measures of the MACS 0416 cluster values to a similar analysis for the the normalization and slope of the galaxy $L-\sigma$ Faber-Jackson relation in the cluster \citep{bergamini2019}. While that analysis used spectral kinematics in a strong lensing context to constrain the $L-\sigma$ relation, our use of weak lensing measurements should provide an independent verification of these fit parameters. While our values for the associated cluster velocity dispersion are measured to higher, we find a great agreement in the measurement of the slope. 
	
	For MACS 0416 they find a Faber-Jackson relation of $L-\sigma^{1/\alpha}$ with $\alpha =$ 0.27 $\pm$ 0.03. Using Equations \ref{eq:massmod} and \ref{eq:Ein_vel} we can produce a Faber-Jackson relation of $\L-\sigma^{2/\ell}$ with a comparative slope of $\frac{\ell}{2} \simeq \alpha$ = 0.25 $\pm$ 0.075. Thus our purely weak-lensing signal is able to return a similar probe of this slope with startling accuracy. It should be noted that while our analyses only overlap in one HFF cluster, their full analysis of three HFF clusters and our two clusters spans four of the HFF program fields and finds consistent agreement in the value of $\frac{\ell}{2} \simeq \alpha$. This finding is accordant with several early-type galaxy cluster population studies \citep{Focardi_2012,KormendyBender}.	
	
	\subsection{Summary}\label{sec:summary}
	In this paper we have developed a parametric approach to constraining cluster properties using only weak-lensing flexion signals. We analyze the two \textit{HST Frontier Fields} clusters, Abell 2744 ($z$ = 0.388) and MACS 0416 ($z$ = 0.397) using cluster member projected flexion. The inclusion of parallel field measurements as well as tangential signals provide a positive proof of radial signal utility in detecting mass in galaxy clusters. The radial flexion decomposition technique provides a strong probe for the ensemble cluster mass using a simple SIS profile, providing a new avenue for weak lensing analysis. 
	
	We have also shown the efficacy of utilizing these second order weak lensing signals to fit the mass-to-light ratio of the member galaxies in large clusters. This is a unique and novel probe of the Faber-Jackson relationship in the first two candidate clusters of the Frontier Fields program and has potential for applications in expanding to the remaining Frontier Fields as well as further cluster studies. 
	
	\section*{Acknowledgements}
	
	This work is based on observations made with the NASA/ESA \textit{Hubble Space Telescope}, obtained from the data archive at the Space Telescope Science Institute, and associated with the Frontier Fields program. STScI is operated by the Association of Universities for Research in Astronomy, Inc. under NASA contract NAS 5-26555. 
	
	This work is based on data and catalog products from HFF-DeepSpace, funded by the National Science Foundation and Space Telescope Science Institute (operated by the Association of Universities for Research in Astronomy, Inc., under NASA contract NAS5-26555) 
	
	This  research made  use  of  \texttt{Astropy},  a  community-developed  core  Python package for Astronomy (Astropy Collaboration et al.  2013),open-source Python modules \texttt{Numpy}, \texttt{Scipy}, \texttt{Matplotlib}, and \texttt{Pandas}. We would like to thank the contributions of Evan J. Arena and Brij Patel in the development of this work. 	We make our full datasets, and analyses code available at \url{https://github.com/DrexelLenser/Lenser}	
	
	\section*{Data Availability}
	
	The data underlying this article are available at https://github.com/DrexelLenser/Lenser.
	
	
	
	
	\bibliographystyle{mnras}
	\bibliography{lensing_MNRAS_2} 

	
	
	

	\bsp	
	\label{lastpage}
\end{document}